\pdfoutput=1
\documentclass[11pt,twoside,a4paper,cmspaper,final,collab]{cms-tdr}
\def\svnVersion{54acca3}\def\svnDate{2025/02/11}\def\cmsCernNoTag{CERN-EP-2024-252}\def\cmsCernDate{\today}\def\cmsMessage{Published in Physics Letters B as \href{http://dx.doi.org/10.1016/j.physletb.2025.139296}{\doi{10.1016/j.physletb.2025.139296}.}}
\begin{document}\cmsNoteHeader{HIG-23-005}

\newcommand{\Irelm}{\ensuremath{I_{\text{rel}}^{\PGm}}}
\newcommand{\Ich}{\ensuremath{I^{\text{ch}}(\mathrm{M})}}
\newcommand{\Ineu}{\ensuremath{I^{\text{neu}}(\mathrm{M})}}
\newcommand{\IchK}{\ensuremath{I^{\text{ch}}(\mathrm{trk}_1)}}
\newcommand{\mllhad}{\ensuremath{m_{\mathrm{M}\PGg}}\xspace}
\DeclareRobustCommand{\PGrz}{{\HepParticle{\PGr}{}{0}}\xspace}
\DeclareRobustCommand{\PGrm}{{\HepParticle{\PGr}{}{-}}\xspace}
\DeclareRobustCommand{\PDstz}{{\HepParticle{\PD}{}{*0}}\xspace} 
\DeclareRobustCommand{\PGrPz}[1]{{\HepParticleResonanceFull{\PGr}{}{}{#1}{}{0}}\Xspace} 
\newcommand{\PGrzM}{\PGrPz{770}}
\DeclareRobustCommand{\PKstzM}{{\HepParticleResonanceFull{\PK}{}{\ast}{892}{}{0}}\xspace} 
\DeclareRobustCommand{\PGfM}{{\PGfP{1020}}}
\DeclareRobustCommand{\PKstarz}{{\HepParticle{\PK}{}{\ast 0}}\xspace}
\DeclareRobustCommand{\PGrm}{{\HepParticle{\PGr}{}{-}}\xspace}
\newcommand{\Hgrho}{\PH\to\PGrz\PGg}
\newcommand{\Hgphi}{\PH\to\PGf\PGg}
\newcommand{\Hgkstar}{\PH\to\PKstarz\PGg}
\newcommand{\gphi}{\PGf\PGg}
\newcommand{\grho}{\PGrz\PGg}
\newcommand{\gkstar}{\PKstarz\PGg}
\newcommand{\ztoll}{\PZ\to\Pell\Pell}
\newcommand{\WH}{\PW\PH}
\newcommand{\ZH}{\PZ\PH}
\newcommand{\ppc}{\Pp\Pp}
\newcommand{\mpipi}{\ensuremath{m_{\PGp\PGp}}}
\newcommand{\pippim}{\PGpp\PGpm}
\newcommand{\kpkm}{\PKp\PKm}
\newcommand{\mkk}{\ensuremath{m_{\PK\PK}}}
\newcommand{\mkpi}{\ensuremath{m_{\PK\PGp}}}
\newcommand{\ggH}{\ensuremath{\Pg\Pg\PH}\xspace}
\newlength\cmsTabSkip\setlength{\cmsTabSkip}{1ex}
\ifthenelse{\boolean{cms@external}}{\providecommand{\cmsTable}[1]{#1}}{\providecommand{\cmsTable}[1]{\resizebox{\textwidth}{!}{#1}}}
\ifthenelse{\boolean{cms@external}}{\providecommand{\cmsLeft}{upper\xspace}}{\providecommand{\cmsLeft}{left\xspace}}
\ifthenelse{\boolean{cms@external}}{\providecommand{\cmsRight}{lower\xspace}}{\providecommand{\cmsRight}{right\xspace}}

\cmsNoteHeader{HIG-23-005}

\title{Search for the Higgs boson decays to a \texorpdfstring{$\PGrz$}{rho}, \texorpdfstring{$\PGf$}{phi}, or \texorpdfstring{$\PKstarz$}{K*0} meson and a photon in proton-proton collisions at \texorpdfstring{$\sqrt{s}=13\TeV$}{sqrt(s) = 13 TeV}}

\date{\today}

\abstract{Three rare decay processes of the Higgs boson to a $\PGrzM$, $\PGfM$, or $\PKstzM$ meson and a photon are searched for using $\sqrt{s}=13\TeV$ proton-proton collision data collected by the CMS experiment at the LHC. Events are selected assuming the mesons decay into a pair of charged pions, a pair of charged kaons, or a charged kaon and pion, respectively. Depending on the Higgs boson production mode, different triggering and reconstruction techniques are adopted. The analyzed data sets correspond to integrated luminosities up to 138\fbinv, depending on the reconstructed final state. After combining various data sets and categories, no significant excess above the background expectations is observed. Upper limits at 95\% confidence level on the Higgs boson branching fractions into $\PGrzM\PGg$, $\PGfM\PGg$, and $\PKstzM\PGg$ are determined to be $3.7 \times 10^{-4}$, $3.0 \times 10^{-4}$, and $3.0 \times 10^{-4}$, respectively. In case of the $\PGrzM\PGg$ and $\PGfM\PGg$ channels, these are the most stringent experimental limits to date.}

\hypersetup{%
pdfauthor={CMS Collaboration},
pdftitle={Search for the Higgs boson decays to a rho, phi, or K(*0) meson and a photon in proton-proton collisions at sqrt(s)=13TeV},%
pdfsubject={CMS},%
pdfkeywords={CMS, Higgs, rare decays}
}

\maketitle

\section{Introduction}
\label{sec:introduction}

In 2012, the ATLAS and CMS experiments at the CERN LHC announced the observation of a new scalar boson~\cite{Aad:2012tfa,Chatrchyan:2012xdj,Chatrchyan:2013lba}. Ever since, a multitude of studies have been performed confirming its compatibility with the standard model (SM) Higgs boson (\PH). In the SM, the Higgs boson interacts with fermion fields via Yukawa-type couplings, which are proportional to the fermion masses. Experimentally, the top quark Yukawa coupling strength is indirectly confirmed through the measurement of the gluon fusion production cross section, as well as by direct measurements of the $\ttbar\PH$ production rate~\cite{Aaboud:2018urx,Sirunyan:2018hoz}. Couplings to the $\PGt$ lepton and \PQb quark are well established thanks to detailed measurements in the respective Higgs boson decay channels~\cite{ATLAS:2022yrq,CMS:2021gxc,Tumasyan_2023,ATLAS:2020fcp,Sirunyan:2018kst}. Interactions with \PQc quarks are searched for via the $\PH$ decay to charm quark jets~\cite{ATLAS:2022ers,CMS:2022psv} or via signatures involving charmonium~\cite{CMS:2022fsq,ATLAS:2022rej,Sirunyan:2018fmm}. Evidence of the Higgs boson coupling to muons has also been established~\cite{CMS:2020xwi}.

Direct measurements of the Higgs boson interaction with low-mass quarks (\PQu, \PQd, \PQs) have not yet been possible because of the small predicted couplings, the challenging reconstruction of the final states, and the overwhelming backgrounds induced by quantum chromodynamic (QCD) processes. In order to probe these couplings, decays of the Higgs boson to a light meson plus either a photon or \PZ boson have been suggested~\cite{Kagan:2014ila}. Upper limits (UL) were obtained for the decays $\PH\to\PZ\PGfM$ and $\PH\to\PZ\PGrzM$~\cite{CMS:2020ggo} by the CMS Collaboration, and for $\PH\to\PGrzM\PGg$, $\PH\to\PGfM\PGg$~\cite{Aaboud:2017xnb}, $\PH\to\PKstzM\PGg$, $\PH\to\PGo(782)\PGg$~\cite{ATLAS:2023alf} and $\PH\to \PDstzP{2010}\PGg$~\cite{atlascollaboration2024searches} by the ATLAS Collaboration.
In the analysis presented here we search for exclusive decays of the Higgs boson to a meson and a photon, where the meson is $\PGrzM$, $\PGfM$, or $\PKstzM$. In what follows we will refer to these mesons as $\PGrz$, $\PGf$, and $\PKstarz$, respectively. For the $\PKstarz$ meson the charge-conjugated partner is implied.

\begin{figure*}[htbp]
  \centering
    \includegraphics[width=0.85\textwidth]{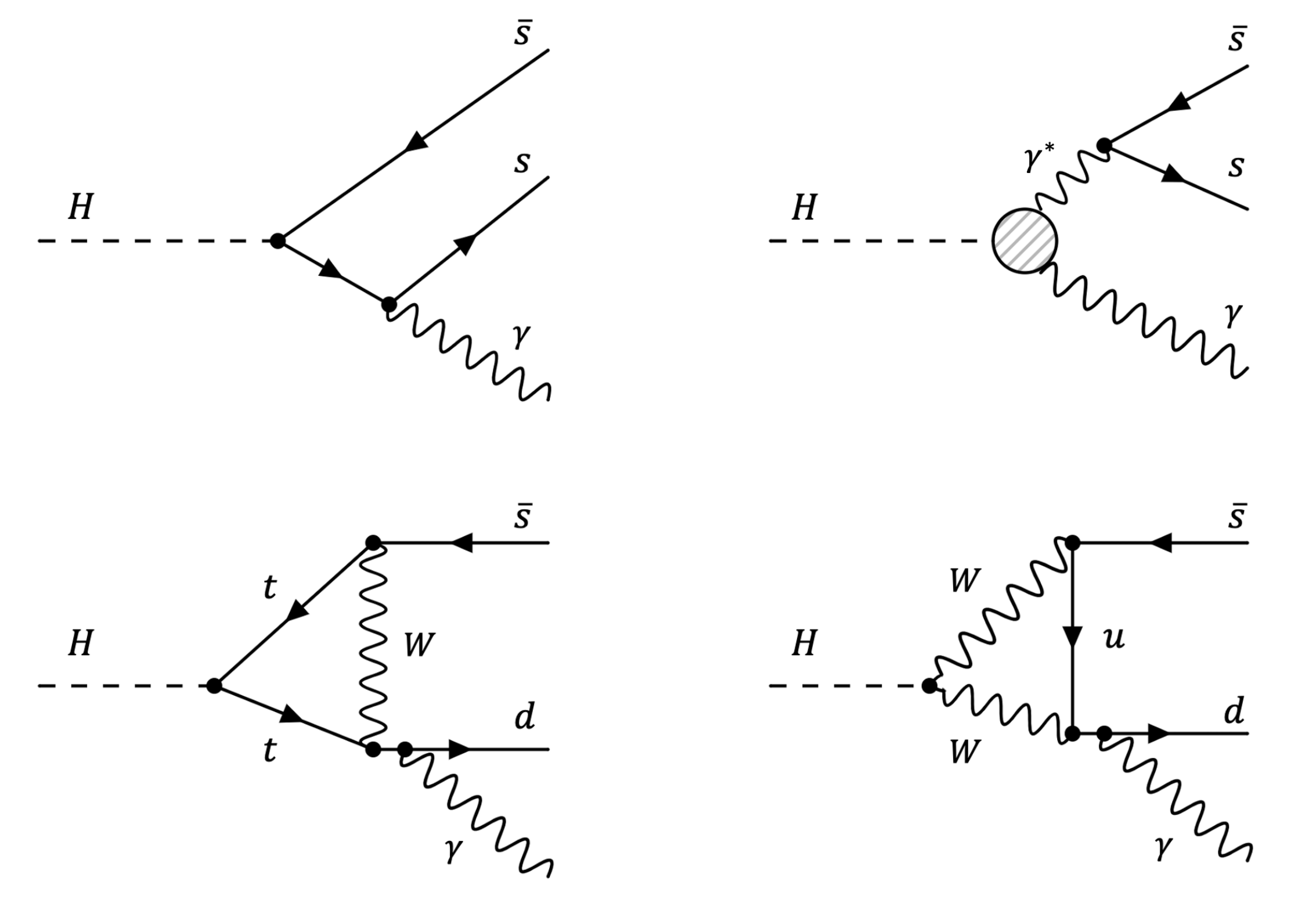}
  \caption{Feynman diagrams showing the different Higgs boson decay mechanisms into a photon and a light meson (upper: $\PGf$ meson; lower: $\PKstarz$ meson). The hatched circle in the upper right diagram denotes the off-shell $\PH \to \PGg\PGg^*$ amplitude, which in the SM arises first at one-loop order.}
  \label{fig:FeynmanDiags}
\end{figure*}

The SM prediction for the branching fraction for $\mathcal{B}(\Hgrho)$ and $\mathcal{B}(\Hgphi)$ are $(1.68\pm0.08)\times10^{-5}$, and $(2.31\pm0.11)\times10^{-6}$, respectively~\cite{Konig:2015qat}. In both decays, the meson is composed of same-flavor quarks as displayed in the upper half of Fig.~\ref{fig:FeynmanDiags} for the $\PGf$ meson. The diagram in the upper left shows the Higgs boson directly decaying to a strange quark-antiquark pair while the diagram in the upper right
displays a Higgs-boson diphoton Dalitz decay with one off-shell photon, which is dominant in the SM.  Beyond-the-SM scenarios with enhanced Higgs boson couplings to light quarks can significantly increase the subdominant contribution and can lead to observable discrepancies that are mostly due to the interference with the dominant diagram~\cite{Kagan:2014ila,Perez:2015lra}.

The $\Hgkstar$ final state features a flavor-changing neutral current and is therefore strongly suppressed by higher electroweak orders in the SM. Example diagrams are shown in the lower half of Fig.~\ref{fig:FeynmanDiags}. The branching fraction for the $\Hgkstar$ decay is $1.0\times10^{-19}$~\cite{dEnterria:2023wjq}, and this decay is therefore a powerful probe of possible anomalous flavor-changing Higgs boson couplings~\cite{Kagan:2014ila}.

For our search, we analyze proton-proton ($\Pp\Pp$) collision data at $\sqrt{s}=13\TeV$, collected with the CMS detector at the LHC. Events are separated into mutually exclusive categories targeting the main production modes of the Higgs boson at the LHC, namely gluon fusion (\ggH), vector boson fusion (VBF), and associated production with a vector boson ($\PV\PH$, $\PV=\PZ,\PW$). The $\PV\PH$ events contain leptonic decays of the vector bosons, $\PW\to\Pell\nu$ and $\PZ\to\Pell\Pell$, where $\Pell=\Pe,\PGm$. The $\PV\PH$ category also includes contributions from events where Higgs boson production occurs in association with top quarks ($\ttbar\PH$), with leptonic decays from the resulting $\PW$ bosons. To reconstruct Higgs boson candidates, photons are combined with meson candidates, identified via their decays $\PGrz \to \PGpp\PGpm$, $\PGf \to \PKp\PKm$ and $\PKstarz \to \PKp\PGpm$.

Tabulated results are provided in the HEPData record for this analysis~\cite{hepdata}.

\section{The CMS detector}
\label{sec:detector}

The central feature of the CMS apparatus is a superconducting solenoid of 6\unit{m} internal diameter, providing a magnetic field of 3.8\unit{T}. Within the solenoid volume are a silicon pixel and strip tracker, a lead tungstate crystal electromagnetic calorimeter (ECAL), and a brass and scintillator hadron calorimeter, each composed of a barrel and two endcap sections. A silicon strip preshower detector improves energy measurements in the forward region. Forward calorimeters extend the pseudorapidity ($\eta$) coverage provided by the barrel and endcap detectors. Muons are reconstructed in gas-ionization detectors embedded in the steel flux-return yoke outside the solenoid.

The silicon tracker measures charged particles within the pseudorapidity range $\abs{\eta} < 2.5$. It consists of 1856 silicon pixel and 15\,148 silicon strip detector modules. The silicon pixel detector modules
are arranged in four layers. In 2016, data were taken with a different detector configuration; at that time there were 1440 silicon pixel detector modules arranged in three layers. The relative \pt resolution for nonisolated particles in the barrel region, $\abs{\eta} < 1.4$, is $\approx1\%$ independent of \pt in the range 1-10\GeV, increases with \pt above 10\GeV, reaching $\approx2\%$ at 50\GeV and $\approx4\%$ at 100\GeV~\cite{TRK-11-001}.

Muons are measured in the pseudorapidity range $\abs{\eta} < 2.4$, with detection planes made using three technologies: drift tubes, cathode strip chambers, and resistive-plate chambers. The single muon trigger efficiency exceeds 90\% over the full $\eta$ range, and the efficiency to reconstruct and identify muons offline is greater than 96\%. Matching muons to tracks measured in the silicon tracker results in a relative \pt resolution, for muons with \pt up to 100\GeV, of 1\% in the barrel and 3\% in the endcaps~\cite{Sirunyan:2018}.

The electron momentum is estimated by combining the energy measurement in the ECAL with the momentum measurement in the tracker. The momentum resolution for electrons with $\pt \approx 45\GeV$ from $\PZ \to \Pe \Pe$ decays ranges from 1.6 to 5\%. It tends to be better in the barrel region than in the endcaps, and also depends on the bremsstrahlung energy emitted by the electron as it traverses the material in front of the ECAL~\cite{CMS:2020uim}.

The energy of photons is obtained from the ECAL energy depositions in a supercluster, combining deposits from the photon and the conversions, produced by the tracker material upstream of the ECAL detector. The energy resolution for photons varies between 8 and 3\% in the \pt range of 20 to 100\GeV~\cite{CMS:2020uim}.

Events of interest are selected using a two-tiered trigger system~\cite{Khachatryan:2016bia}. The first level, composed of custom hardware processors, uses information from the calorimeters and muon detectors to select events at a rate of around 100\unit{kHz} within a fixed latency of 4\mus~\cite{CMS:2020cmk}. The second level, known as the high-level trigger, consists of a farm of processors running a version of the full event reconstruction software optimized for fast processing, and reduces the event rate to around 1\unit{kHz} before data storage~\cite{CMS:2024aqx}.

A more detailed description of the CMS detector, together with a definition of the coordinate system used and the relevant kinematic variables, can be found in Ref.~\cite{Chatrchyan:2008zzk}.

\section{Simulated samples}
\label{sec:signal-simulation}

Samples of simulated Higgs boson events, produced via \ggH, VBF, $\ttbar\PH$, $\WH$, and $\ZH$, are generated at next-to-leading order (NLO) in QCD using \POWHEG 2.0, with improved accuracy where available~\cite{Nason:2004rx,Frixione:2007vw,Alioli:2008tz,Alioli:2010xd,Nason:2009ai,Luisoni:2013kna}. The signal yields are normalized to the theoretical cross sections. The \ggH cross section is calculated at next-to-next-to-NLO in QCD and at NLO in electroweak accuracy, while those for the other production modes are calculated at next-to-NLO in QCD and at NLO in electroweak accuracy~\cite{deFlorian:2016spz}. The decay $\PGrz\to\pippim$ has a branching fraction of $\approx100\%$, while the decay $\PGf\to\kpkm$ has a branching fraction of $(49.1 \pm 0.5)\%$. For the decay $\PKstarz\to\PK\PGp$, the branching fraction is close to $100\%$, with two-thirds of these decays involving a charged kaon and a charged pion~\cite{piddigi}.

Simulated samples are utilized to measure data-to-simulation correction factors for the object identification and trigger selection efficiency; these simulations are also employed to optimize the event selections within each category. The related SM processes are: $\PGg$, $\PW\to\Pell\nu$, and Drell--Yan $\ztoll$ production in association with jets, as well as $\ttbar$, $\PW\PGg$, and $\PZ\PGg$ production. These samples are generated at leading order (LO) using \MGvATNLO 2.6.5 with MLM parton matching~\cite{Alwall:2014hca}, \MGvATNLO 2.6.5 at NLO, with FxFx matching~\cite{Frederix:2012ps} or \POWHEG program. The NNPDF3.1 parton distribution functions (PDFs)~\cite{Ball:2017nwa} at next-to-NLO order are used for all samples. 

All generated samples are interfaced with \PYTHIA 8.212~\cite{Sjostrand:2014zea} to model parton showering and hadronization. The description of the underlying event is provided by the CP5 tune~\cite{Sirunyan:2019dfx}. For the signal samples the decays $\Hgrho$, $\Hgphi$, or $\Hgkstar$ are also modeled using \PYTHIA. 
With the Higgs boson being a scalar particle, angular momentum conservation enforces fully transverse spin alignment of the outgoing vector meson, but \PYTHIA simulates unpolarized decay products. To account for this effect, signal events are weighted at generator level by a factor proportional to $\sin^2\theta$, where $\theta$ is the angle between the direction of the positive kaon or pion in the meson rest frame and a polarization axis defined as the meson flight direction in the Higgs boson rest frame; the polarization axis can be effectively approximated, in the present case, with the meson flight direction measured in the laboratory, given the smallness of the meson masses with respect to the Higgs boson mass~\cite{PPHEP}.
A validation of this reweighting has been performed using \EVTGEN~\cite{LANGE2001152}.

Additional inelastic $\ppc$ collisions in each bunch crossing, referred to as pileup (PU), are generated with \PYTHIA and are added to all simulated events in accordance with the measured PU distribution. All generated events are processed through a \GEANTfour-based~\cite{Agostinelli:2002hh} simulation of the CMS detector before being reconstructed with the same version of the CMS event reconstruction software used for data. Detector conditions corresponding to the different data-taking periods in years 2016, 2017, and 2018 are simulated separately.

\section{Event reconstruction}
\label{sec:event-reconstruction}

The final-state particles of $\ppc$ collisions are reconstructed with a particle-flow (PF) algorithm~\cite{Sirunyan:2017ulk}, which combines information from all subdetectors to reconstruct individual particle candidates (PF candidates). These particle candidates are classified as muons, electrons, photons, charged and neutral hadrons.

The primary vertex (PV) is taken to be the vertex corresponding to the hardest scattering in the event, evaluated using tracking information alone, as described in Section 9.4.1 of Ref.~\cite{CMS-TDR-15-02}. All other collisions in the event are considered to have originated from PU. The candidate muons, electrons, and charged-hadron tracks used in the search presented in this paper are all required to originate from the PV.

Candidate photons from the particle-flow algorithm must satisfy a tight (medium) identification multivariate analysis (MVA) criterion corresponding to a true-photon efficiency of 80\% (90\%)~\cite{CMS:2020uim}, depending on the event selection described in Section~\ref{sec:event-selection}. Discriminating variables include photon kinematic observables and ECAL-to-preshower energy ratios in the endcaps, variables related to electromagnetic shower shape and width, and the isolation of the photon candidate. The isolation variable is computed either considering only tracks or all PF candidates in a cone within $\Delta R=0.4$ from the cluster centroid, where $\Delta R = \sqrt{\smash[b]{(\Delta\eta)^{2} + (\Delta\phi)^{2}}}$ and $\Delta\eta$ and $\Delta\phi$ are differences in pseudorapidity and azimuthal angle, respectively. Vetoes are applied in order to reject electron candidates and photons that converted into an electron-positron pair~\cite{CMS:2020uim}. Corrections on photon energy scale, resolution, and identification efficiency, derived from $\PZ\to\Pe\Pe$ events~\cite{CMS:2020uim} using the ``tag-and-probe'' (TnP) method~\cite{Khachatryan:2010xn}, are applied to simulated events.

Muons are required to satisfy identification and isolation criteria with an average selection efficiency of about 95\%. An isolation variable ($I^{\PGm}_\text{rel}$) for muons is determined by summing the $\pt$ of all PF candidates within a cone of radius $\Delta R = 0.4$ around the direction of the candidate muon and is corrected for the contribution of neutral particles from PU interactions~\cite{Cacciari:2008gn}. Factors correcting for the identification efficiency of muons have been computed using the TnP method.

To reduce contamination from incorrectly reconstructed particles, candidate electrons are required to pass a tight (medium) MVA electron identification discriminant corresponding to an efficiency of 80\% (90\%), depending on the event selection. This discriminant~\cite{CMS:2020uim} combines information about the quality of the tracks, the shower shape, kinematic quantities, and hadronic activity in the vicinity of the reconstructed electron. Electron energy scale and resolution corrections, derived from $\PZ\to\Pe\Pe$ events~\cite{CMS:2020uim}, are applied to simulated events. The TnP method is also used to determine correction factors of the electron identification efficiency~\cite{CMS:2020uim,Sirunyan:2018}.

Jets are reconstructed from PF candidates using the anti-$\kt$ clustering algorithm~\cite{Cacciari:2008gp,Cacciari:2011ma}
with a distance parameter of $0.4$. To mitigate the effect of PU, charged hadrons that do not arise from the PV are removed from the clustering. In addition to jet identification criteria, PU rejection criteria are used to reduce the contamination of jets with $\pt^\mathrm{j}<50\GeV$ initiated by PU interactions~\cite{CMS:2020ebo}. Differences in jet energy scale and resolution between data and simulation are corrected for, as described in Ref.~\cite{CMS:2016lmd}.

The meson candidates are identified using tracks reconstructed in the tracking detector. The tracks are required to originate from the PV and to satisfy the ``high purity'' reconstruction requirements. These requirements are based on the number of tracker layers with reconstructed hits, the track fit quality, and the values of the impact parameters relative to their uncertainties~\cite{TRK-11-001}. A $\PGrz$, $\PGf$\, or $\PKstarz$ meson candidate, denoted generically as ``M'', is reconstructed as a pair of oppositely-charged tracks, one with $\pt>20\GeV$ and the other with $\pt>5\GeV$; both within $\abs\eta<2.5$. The pseudorapidity of the meson is restricted to match the trigger requirements, described in Section~\ref{sec:event-selection}. The meson decay vertex is determined with a kinematic vertex-constrained fit~\cite{Chabanat:2005wj} and track momenta are recalculated accordingly. An isolation variable of the meson candidate is determined using PF candidate transverse momenta as:
\ifthenelse{\boolean{cms@external}}
{ 
\begin{equation}
  \label{eq:isolations}
  \begin{aligned}
  \Ich &= \frac{ p_{\mathrm{T}}^\mathrm{M} }{ p_{\mathrm{T}}^\mathrm{M} + \sum_\text{ch} \lvert p_{\mathrm{T}}^\text{ch}\rvert },\\
  \Ineu &= \frac{ p_{\mathrm{T}}^\mathrm{M} }{ p_{\mathrm{T}}^\mathrm{M} + \sum_\text{neu} \lvert p_{\mathrm{T}}^\text{neu}\rvert },
  \end{aligned}
\end{equation}
} 
{ 
\begin{equation}
\label{eq:isolations}
  \Ich = \frac{ p_{\mathrm{T}}^\mathrm{M} }{ p_{\mathrm{T}}^\mathrm{M} + \sum_\text{ch} \lvert p_{\mathrm{T}}^\text{ch}\rvert }
  \quad\text{and}\quad
  \Ineu = \frac{ p_{\mathrm{T}}^\mathrm{M} }{ p_{\mathrm{T}}^\mathrm{M} + \sum_\text{neu} \lvert p_{\mathrm{T}}^\text{neu} \rvert },
\end{equation}
} 
where $\sum_\text{ch} \abs{p_{\mathrm{T}}^\text{ch}}$ and $\sum_\text{neu} \abs{p_{\mathrm{T}}^\text{neu}}$ are, respectively, the \pt sums of tracks and neutral particle candidates in a cone of radius $\Delta R = 0.3$ around the direction of the two-track system. Only tracks with $\pt > 0.9\GeV$ associated with the meson vertex, excluding the candidates defining the meson, and only neutral candidates with $\pt > 0.5\GeV$, are retained in the isolation variable calculation. Data-to-simulation correction factors accounting for differences in track- and neutral particles based isolation variables are derived from events with two oppositely-charged muons using the following TnP method. After identifying a high-quality muon, isolation variables are computed from the activity around a second candidate muon in the event, satisfying looser criteria, which serves as a proxy for the candidate meson. The efficiency of the isolation requirement is extracted fom a simultaneous $\PZ+$background fit to the dimuon mass distribution for events passing or failing the isolation requirement. This efficiency is compared to the ditrack isolation efficiency estimated from the signal samples to derive corrections, which amount to values around 1\% depending on $\pt^\PGm$. A systematic uncertainty arising from using a muon as a proxy for a meson is applied and detailed in Section~\ref{sec:uncertainties-inference}.

\begin{figure*}[htbp]
\centering
\includegraphics[width=0.49\textwidth]{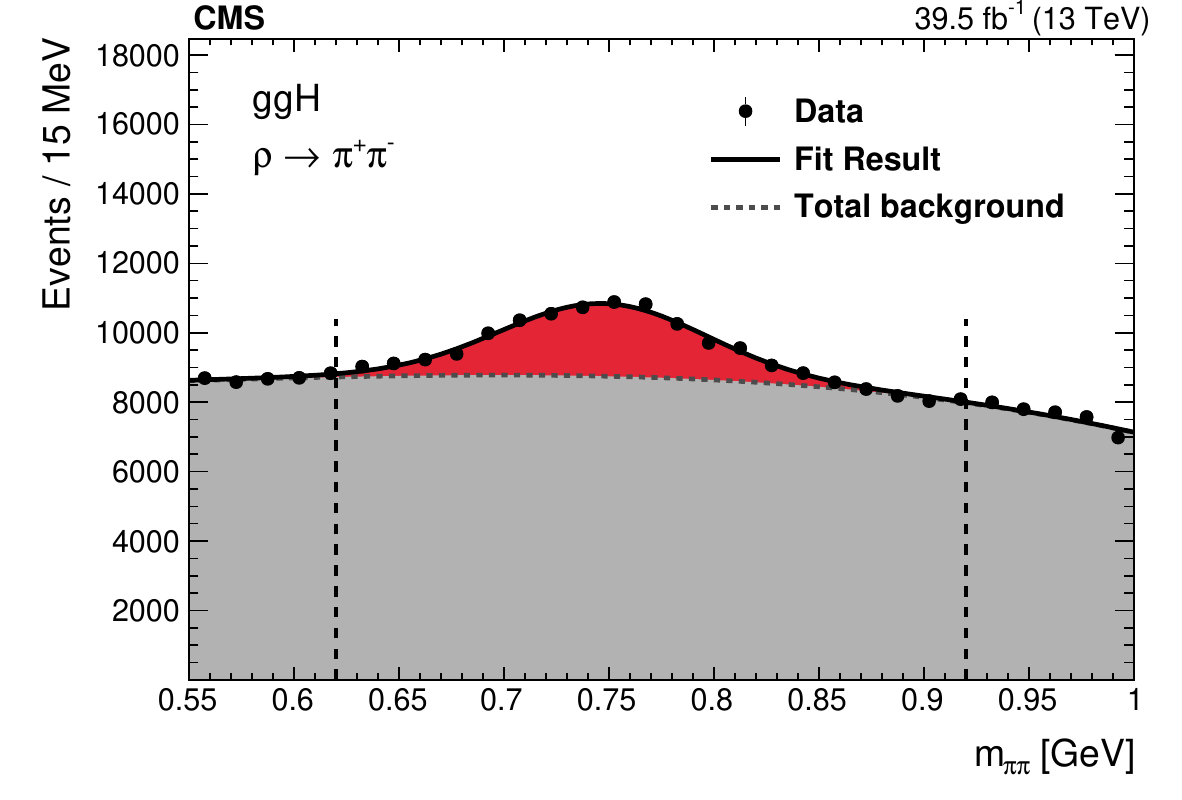}
\includegraphics[width=0.49\textwidth]{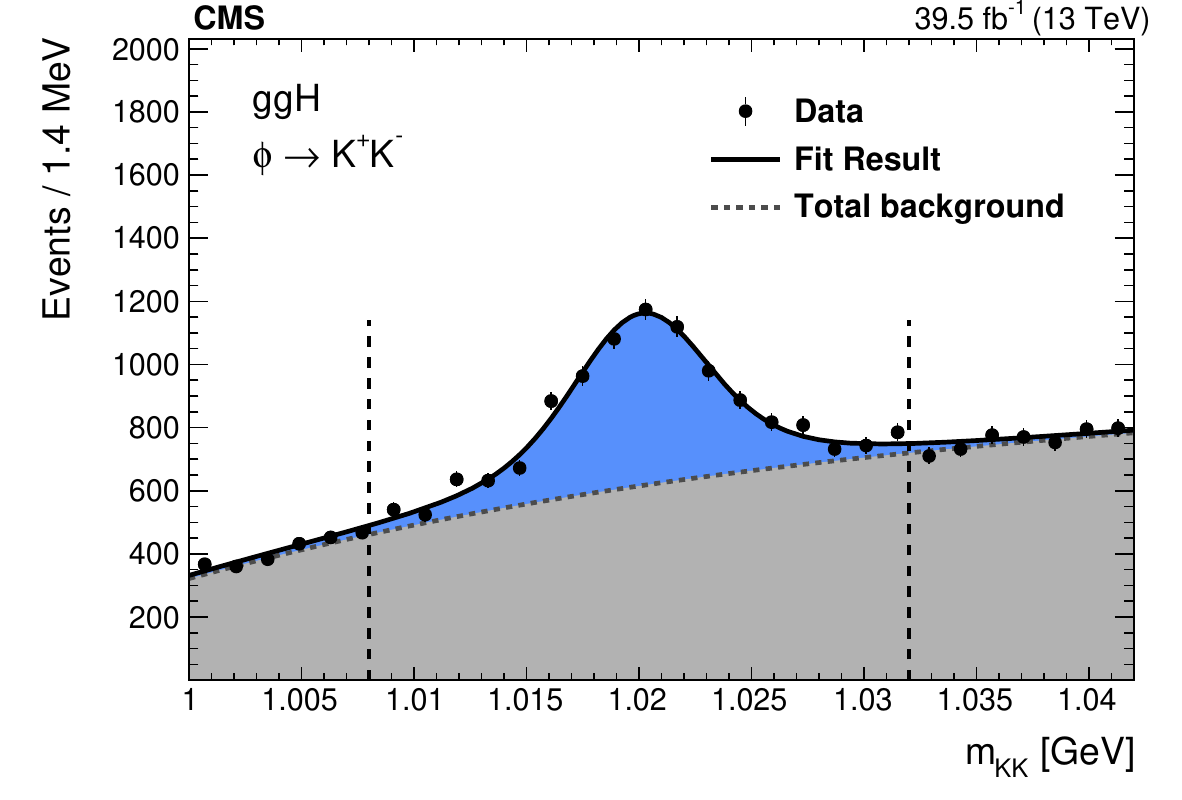}\\
\includegraphics[width=0.49\textwidth]{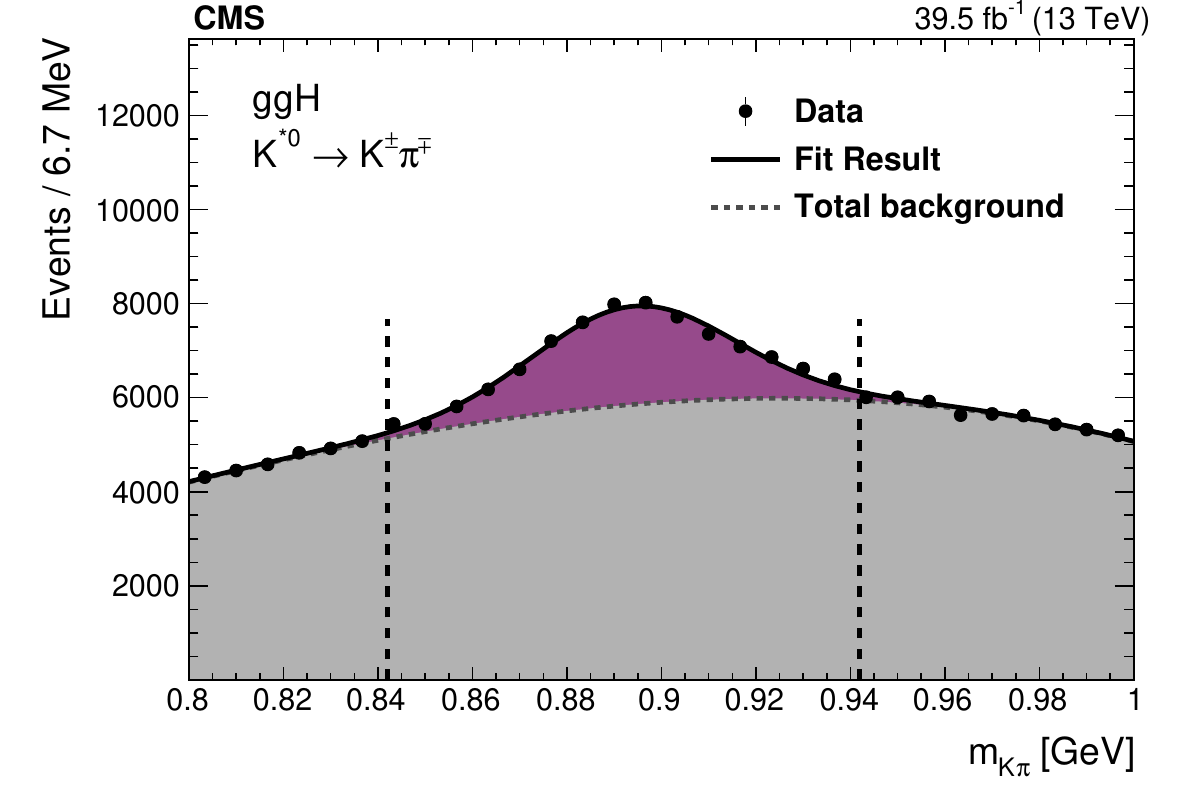}
\caption{Track pair invariant mass distributions in selected data events, for the \ggH category of the analysis, for the $\grho$ (upper left), $\gphi$ (upper right), and $\gkstar$ (lower) decays. The vertical dashed lines represent the signal mass region borders.}
\label{fig:ggfplots}
\end{figure*}

It is required that the track-pair mass, under the appropriate charged-pion or charged-kaon mass hypotheses for the tracks, is consistent with the corresponding nominal meson mass. The experimental resolution ranges from approximately $3\unit{\MeV}$ for {\mkk} to $8\unit{\MeV}$ for {\mpipi}. Events are selected if the mass of the ditrack system is within the ranges defined in Table~\ref{table:mditrk_regions}. In the $\PKstarz$ case, if both the $\PKp\PGpm$ and the $\PKm\PGpp$ candidates fall in the mass interval, the one with the mass closest to the nominal $\PKstarz$ mass is chosen. No attempt is made to make these selections mutually exclusive. Moreover, if more than one meson candidate in the relevant mass intervals are found in the same event, only the one with the largest $\pt$ is retained in the corresponding search. The mass intervals located on either side of the signal mass regions form a \emph{sideband region}, as defined in Table~\ref{table:mditrk_regions}. The sideband region is used for training an MVA discriminator in the \ggH category, and to validate another MVA discriminator used in the VBF category, as described in Section~\ref{sec:mva}. The distributions of the ditrack system mass in the three decay channels are shown in Fig.~\ref{fig:ggfplots}, exhibiting a visible resonance peak at the mass of the meson. The meson peaks are fitted with a Voigtian function and the background with a third-order Chebychev polynomial. Figure~\ref{fig:ggfplots} provides a qualitative illustration of the meson selection in the \ggH category.

\begin{table*}[htbp]
  \centering
  \topcaption{Mass requirements imposed on the ditrack system for signal and sideband regions.}
  \renewcommand{\arraystretch}{1.3}
  \begin{tabular}{l c c c}{}
  meson decay & lower boundary ($\GeVns$) & signal region ($\GeVns$) & upper boundary ($\GeVns$) \\
  \hline
  $\PGrz\to\pippim$  & $0.55$ & $0.62$ -- $0.92$ & $1.00$ \\
  $\PGf\to\kpkm$  & $1.000$ & $1.008$ -- $1.032$ & $1.042$ \\
  $\PKstarz\to\PKp\PGpm$ & $0.80$ & $0.84$ -- $0.94$ & $1.00$ \\
  \end{tabular}
\label{table:mditrk_regions}
\end{table*}

\section{Event selection}
\label{sec:event-selection}

In the following, four analyses categories that target $\ttbar\PH$, $\PV\PH$, VBF, and \ggH processes are discussed. The corresponding event selection criteria are summarized in Table~\ref{tab:allsel}. In the \ggH and VBF categories, the background is mostly composed of $\PGg$+jet and multijet events, which occur at rates many orders of magnitude larger than that of Higgs boson production.
Main backgrounds in the $\PV\PH$ category are $\PW\to\Pell\nu$, and Drell--Yan $\ztoll$ production in association with jets, as well as $\ttbar$, $\PW\PGg$, and $\PZ\PGg$ events.

\subsection{\texorpdfstring{$\PV\PH$}{VH} category}
The $\PV\PH$ category targets production of a Higgs boson in association with a heavy vector boson. However signal events selected under $\PV\PH$ category also include $\ttbar\PH$ events where one or both $\PW$ bosons from top quark decays yield leptons, accounting for approximately 30\% of the event statistics in this category. Trigger selection is based on single- or double-lepton requirements, where the single-muon (electron) lowest \pt thresholds vary between 24 and 27\GeV (27 and 32\GeV), depending on the data-taking period, while the double-muon (electron) lowest \pt thresholds are 17 and $8\GeV$ (23 and $12\GeV$). Triggers requiring a muon with $\pt^\PGm>17\GeV$ and a photon with $\pt^\PGg>30\GeV$ complement the selection. These triggers were active for the full data-taking periods in years 2016, 2017, and 2018 corresponding to 138 \fbinv.
 
In the $\PW\to\Pell\nu$ decay channel, events are retained if they contain a muon with $\pt^\PGm > 20\GeV$ and $\Irelm<0.15$, or an electron with $\pt^\Pe > 30\GeV$, where both must satisfy tight identification criteria. In the $\PZ\to\Pell\Pell$ decay channel, two muons are required with $\pt^\PGm > 20$ and $10\GeV$, and $\Irelm<0.25$, or two electrons with $\pt^\Pe > 25$ and 15\GeV, where both must satisfy tight identification criteria. All electrons (muons) must be within $\abs{\eta}<2.5\,(2.4)$. The ZH category includes events with two opposite-charge and same-flavor leptons if they satisfy $\abs{M_{\Pell\Pell}-M_\PZ} < 15\GeV$.

Events are required to contain photons with $\pt^\PGg > 40\GeV$, $\abs{\eta^\PGg}< 2.5$ and satisfying medium identification criteria. Meson candidates must have $\pt^{\mathrm{M}}>40\GeV$ and $\Ich > 0.8$. In the single-muon final state, it is required that $\Delta R(\PGm,M)>0.5$. In the single-electron final state, events with electron-photon pairs consistent with the \PZ boson decays are vetoed by requiring $|M_{\Pe\PGg}-M_\PZ|>10\GeV$ to reduce background from electrons misidentified as photons. In the dilepton final state, an additional requirement is imposed on the mass of the dilepton-$\PGg$ system, $M_{\Pell\Pell\PGg}>96\GeV$ in order to reduce the impact of final-state radiation (FSR).
A lepton veto is applied in the other categories described below to ensure they are mutually exclusive.

\subsection{High-\texorpdfstring{$\pt^\PGg$}{pt gamma} VBF category}
The high-$\pt^\PGg$ VBF category targets VBF events with a high-energy photon. During the data taking these events are selected by a trigger requiring a photon with $\pt^\PGg > 75\GeV$, in the ECAL barrel region ($\abs{\eta^\PGg}< 1.4$), and a pair of jets with large invariant mass ($M_{\mathrm{jj}}>400\GeV$) and large separation in pseudorapidity ($\abs{\Delta\eta_{\mathrm{jj}}}>3$). The threshold values were adjusted throughout the data-taking years. This trigger was used for part of 2016 (28.2 \fbinv) and part of 2017 (7.7 \fbinv), and for all of 2018 (63.7 \fbinv). 

Events are required to contain photons with $\pt^\PGg > 75\GeV$, that satisfy medium photon identification criteria. Photons in the ECAL endcaps are not considered because the trigger was limited to the ECAL barrel region.
Events must also include meson candidates with $\Ich > 0.9$ and $\pt^{\mathrm{M}}>30\GeV$.  
In addition, two jets not overlapping with the $\PH\to\mathrm{M}\PGg$ signal candidate are required, which must satisfy $\pt^\mathrm{j}> 40\GeV$, $M_\mathrm{jj} > 400\GeV$, and $\abs{\Delta\eta_{\mathrm{jj}}} > 3$.

\subsection{Low-\texorpdfstring{$\pt^\PGg$}{pt gamma} VBF category}
Since the VBF trigger is ineffective for low-$\pt$ photons, we adopt a trigger strategy introduced in 2018 (39.5\fbinv), only based on the Higgs decay products. This trigger selects a photon and a jet each with $\pt > 35\GeV$. In addition, the jet must be compatible with a single, isolated ditrack system. The photon selection is based on the ratio of the HCAL and ECAL cluster energies, shower-shape parameters, and isolation computed from ECAL and HCAL energy deposits and tracker information, in a cone of $\Delta R = 0.4$ around the candidate photon. Utilizing the similarity among the tau lepton hadronic decay and the meson signature, trigger requirements on the ditrack system follow those of $\PGt$ reconstruction. Final states with two tracks and no \PGpz mesons are selected requiring the tight working point for \PGt identification and isolation~\cite{CMS:2018jrd}. Finally the photon-jet system must satisfy $\Delta R_{\mathrm{j},\PGg} > 0.3$ and $M_{\mathrm{j},\PGg} > 30\GeV$. The trigger efficiency for the Higgs boson signal is around 55\% with respect to the offline selection.

The efficiency of the photon and $\PGt$-like jet reconstruction at trigger level was studied by factorizing the trigger efficiency into the photon and $\PGt$-like jet parts, and using two dedicated triggers.
These triggers require a muon with $\pt^\PGm > 24\GeV$, as well as a photon or a $\PGt$-like jet with the same requirements as the signal. In data and simulated events, a $\PZ\to\PGm\PGm$ boson candidate and a photon or a meson candidate are reconstructed,  
where one of the two candidate muons from the \PZ boson matches the muon trigger objects. The meson (photon) trigger efficiency for data and simulation is determined by measuring the fraction of the $\PZ$-plus-meson (-photon) events that also pass the dedicated trigger requirement. The photon trigger efficiency is determined using events where the photon-plus-dimuon mass is compatible with that of the \PZ boson.

Selected events are required to contain at least one photon with $38 < \pt^\PGg < 75\GeV$, $\abs{\eta^\PGg}< 2.1$ that passes the medium (tight) identification criteria in the ECAL barrel (endcap). Events are required to have at least one meson candidate with $\pt^{\mathrm{M}}>38\GeV$ and $\Ich > 0.9$. To confirm the VBF nature of the category, we require a pair of jets with $\pt^\mathrm{j}> 30$ and $20\GeV$, $M_\mathrm{jj} > 300\GeV$, and $\abs{\Delta\eta_{\mathrm{jj}}} > 3$, where the jets do not overlap with the $\PH\to\mathrm{M}\PGg$ signal candidates. In the following, jets satisfying these criteria are referred to as VBF-like jets.

\subsection{\texorpdfstring{\ggH}{ggH} category}
The \ggH category includes events selected by the same trigger as the low-$\pt^\PGg$ VBF category. After a VBF-like jets veto is applied and the overlap with other categories is removed, events are required to contain photons with $\pt^\PGg > 38\GeV$, $\abs{\eta^\PGg}< 2.5$ and passing the tight identification criteria. Since the mesons from the Higgs boson decays have a large momentum, their secondary decay products are expected to be collimated, and hence have a jet-like signature. As meson candidate tracks, we consider the two leading-\pt charged constituents of a jet which satisfies basic jet requirements~\cite{CMS-PAS-JME-10-003}, and that lie in a cone with $\Delta R = 0.07$ between the tracks direction, thus strongly reducing track combinations. Candidate mesons with $\pt^{\mathrm{M}}>38\GeV$, $\Ich > 0.9$ and $\Ineu > 0.8$ are retained. This category includes approximately 4\% of VBF signal events.

\section{Multivariate analysis selection}
\label{sec:mva}

An analysis-specific MVA selection, based on a boosted decision tree (BDT) algorithm, has been developed to improve the signal-to-background ratio in the \ggH, VBF low-$\pt^\PGg$, and VBF high-$\pt^\PGg$ categories. The BDT classifiers use gradient boosting~\cite{XGBoost} implemented in the \textsc{Root} package TMVA~\cite{TMVA}. These classifiers are trained on half of the signal and background events, and validated with the other half. Choice of the BDT input variables aims to achieve a robust signal-to-background
discrimination ensuring low inter-variable correlation.

In the \ggH category, simulated signal and data events from the meson mass sidebands defined in Section~\ref{sec:event-reconstruction} are used as training and test samples. The BDT is based on four variables: the ratio of the meson \pt to the photon and ditrack three-body mass (\mllhad), $\pt^\mathrm{M}/\mllhad$; the ratio of the photon \pt and \mllhad, $\pt^\PGg/\mllhad$; the meson pseudorapidity; and the isolation of the leading meson track candidate,
\begin{equation}
\IchK = \frac{p_{\mathrm{T}}^{\text{trk}_1}}{ p_{\mathrm{T}}^{\text{trk}_1} + \sum_\text{ch} \lvert p_{\mathrm{T}}^\text{ch} \rvert },
\end{equation}
where $\sum_\text{ch} \abs{p_{\mathrm{T}}^\text{ch}}$ is defined as in Eq. (\ref{eq:isolations}).
Regarding the ratio variables, the photon and meson \pt are divided by \mllhad\ in order to decrease their correlation with the Higgs boson mass estimate.

In the VBF categories, background from simulated photon plus jet events and photon plus ditrack data events are used for training and testing. In the data events, the two tracks are required to have the same charge. The list of BDT variables used is as follows:
$\pt^\mathrm{M}/\mllhad$; $\pt^\PGg$;  
$\pt^{\mathrm{M}\PGg}$; the photon identification discriminator; the meson charged isolation; $\mathrm{M}_{\mathrm{jj}}$; the $\Delta\phi$ of the two tagging jets; and the Zeppenfeld variable, defined as $z^* = \abs{\eta_{\mathrm{M}\PGg}-0.5(\eta_{\mathrm{j}1}+\eta_{\mathrm{j}2})}/\abs{\Delta\eta_{\mathrm{jj}}}$~\cite{Rainwater:1996ud}.  

Each event category employing a BDT-based classification is further split in two sub-categories of different purity, cat0 and cat1, where the BDT score thresholds of the two sub-categories have been optimized for a maximal value of $\varepsilon/\sqrt{B}$. Here $\varepsilon$ and $B$ denote the signal efficiency and number of expected events of the \mllhad distribution, within a range of $\pm 3$ standard deviations centered around the Higgs boson mass, for the given BDT score selection. 
In each case, the two sub-categories are analyzed separately for signal extraction. Events with a BDT-score below the lowest threshold (cat1) are discarded. Table~\ref{tab:allsel} summarizes selection criteria for all event categories. 

\begin{table*}[htbp]
\centering
\topcaption{Summary of the event selection criteria, photon identification efficiency and integrated luminosity ($\mathcal{L}$) used in the analysis.}
\renewcommand{\arraystretch}{1.3}
\cmsTable{
\begin{tabular}{lcccc}
\multicolumn{5}{c}{Common selections} \\
\hline
& \multicolumn{4}{c}{2 ``high-purity'' tracks, opposite charge} \\
M selection & \multicolumn{4}{c}{$\abs{\eta^{\text{trk}}}<2.5$, $\pt^{\text{trk}1}> 20\GeV$, $\pt^{\text{trk}2}> 5\GeV$, $\abs{\eta^{\mathrm{M}}}<2.1$} \\
& \multicolumn{4}{c}{$0.62<\mpipi<0.92\GeV$ (\PGrz) / $1.008<\mkk<1.032\GeV$ (\PGf) / $0.84<\mkpi<0.94\GeV$ (\PKstarz) } \\
\hline
\multicolumn{5}{c}{} \\
Category & \ggH & VBF High-$\pt^\PGg$ & VBF Low-$\pt^\PGg$ & $\PV\PH$ \\
\hline
$\mathcal{L}$ & $39.5\fbinv$ & $86.9\fbinv$ & $39.5\fbinv$ & $138\fbinv$\\
Trigger & $\PGg$ $+$ & High-$\pt$ $\PGg$ $+$ &  $\PGg$ $+$ & Double or single $\Pe$/$\PGm$\\
& jet with $\PGt$-ID & VBF-like jets & jet with $\PGt$-ID & \\
$\pt^\PGg$ (\GeVns{}) & $>$38 & $>$75 & $>$38 and $<$75 & $>$40 \\
$\abs{\eta^\PGg}$ & $<$2.5 & $<$1.4 & $<$2.1 & $<$2.5 \\
$\PGg$-ID (eff.) & 80\% & 90\% & 80\% & 90\% \\
$\pt^\mathrm{M}$ (\GeVns{}) & $>$38 & $>$30 & $>$38 & $>$40 \\
$\Ich$ & $>$0.9 & $>$0.9 & $>$0.9 & $>$0.8 \\
$\Ineu$ & $>$0.8 & \NA  & \NA  & \NA  \\
\hline
Event & Meson candidate & 2 jets with & 2 jets with & 1 selected and \\
tagging & within a jet with & $\pt^\mathrm{j} > 40\GeV$ & $\pt^\mathrm{j} > 30/20\GeV$ &  isolated $\Pe$/$\PGm$ \\ 
& $\pt^\mathrm{j} > 40\GeV$, tracks & $m_\mathrm{jj} > 400\GeV$ & $m_\mathrm{jj} > 300\GeV$ & or 2 selected $\Pe$/$\PGm$ \\
& with $\Delta R < 0.07$ & $\abs{\Delta\eta_\mathrm{jj}} > 3$ & $\abs{\Delta\eta_\mathrm{jj}} > 3$ & $\abs{M_{\Pell\Pell}-M_\PZ} < 15\GeV$ \\
Veto & $\Pe$/$\PGm$, VBF-like jets & $\Pe$/$\PGm$ & $\Pe$/$\PGm$ & \NA \\
\hline
\multicolumn{5}{c}{BDT categories} \\
\hline
cat0 & BDT $> 0.55$ & BDT $> 0.7$ & BDT $> 0.7$ & \NA  \\ 
cat1 & $-0.4 < $ BDT $< 0.55$ & $-0.6 < $ BDT $ < 0.7$ & $-0.6 < $ BDT $ < 0.7$ & \NA  \\ 
\hline 
\end{tabular}
}
\label{tab:allsel}
\end{table*}

\section{Modeling of the \texorpdfstring{\PH}{H} candidate mass distribution and analysis strategy}
\label{sec:background-modeling}

The $\mllhad$ distributions are used in the signal extraction. If multiple photon and ditrack candidates are selected in an event, only the one with the largest \pt is retained to form the Higgs boson candidate. The analysis categories and the three data-taking periods, corresponding to the years 2016, 2017, and 2018, are considered as separate contributions in a simultaneous fit.

In each category and decay channel, the $\mllhad$ signal distributions are described by analytical functions with their parameters determined through unbinned maximum likelihood fits to the respective samples. Based on the simulated signal events, the shape of the $\mllhad$ distribution is approximated by a double-sided Crystal Ball function (DSCB)~\cite{Oreglia:1980cs}. The values of the DSCB parameters, extracted from fits to simulated signals in different categories and decay channels, are fixed in the final fits to data.

The background parametrization is determined with fits in the $\mllhad$ spectrum sidebands $m_{L} < \mllhad < 120\GeV$ and $130\GeV < \mllhad < m_{R}$, where $\{m_L, m_R\} = \{100, 150\}\GeV$ for the leptonic decay, $\{m_L, m_R\} = \{100, 170\}\GeV$ for the high-$\pt^\PGg$ VBF category, and $\{m_L, m_R\} = \{110, 160\}\GeV$ for the remaining ones. We use Chebychev and Bernstein polynomials, along with exponential functions to model the background shape.

The optimal order of polynomial functions, that vary between two and four depending on the category and decay channel, is determined using a Fisher statistical test~\cite{Fisher:1922}. All parameters of the functions are allowed to float freely in the final fits to data. The potential bias due to the choice of the background parametrization is evaluated by generating pseudo-experiments following the distribution of one of the background models. Amounts of signal corresponding to branching fractions of $10^{-5}$ are also added to pseudo-data. Extended unbinned maximum likelihood fits, using an alternative background model, are performed to extract the branching fraction from each pseudo-experiment. The median difference between the measured and injected signal yields, relative to the post-fit uncertainty in the signal yields, gives an estimate of the bias induced by the choice of the fitting function as background model. The measured biases are found to be smaller than 20\% across all background parametrizations. Including these observed deviations as spurious signals leads to a negligible change in the overall uncertainty of the measured signal rate of less than 1\%. For the final systematic uncertainty estimation related to the background shape, including a possible bias, all tested background parametrizations are employed in a discrete profiling method~\cite{Dauncey_2015} described in the following.

\section{Systematic uncertainties}
\label{sec:uncertainties-inference}

To estimate systematic uncertainties related to the choice of the $\mllhad$ analytical background shapes, a discrete profiling method is used, which treats the choice of a specific model as a discrete nuisance parameter in the fit. Several systematic uncertainties affect the normalization of the simulated signal templates; they are included in the likelihood function as nuisance parameters and are profiled when minimizing it.
\renewcommand{\theenumi}{\roman{enumi}}
\begin{enumerate}
\item The uncertainties in the integrated luminosity measurements range between 1.2 and 2.5\%, depending on data-taking period~\cite{CMS:2021xjt,CMS:2018elu,CMS:2019jhq}. 
\item The uncertainty in the total inelastic cross section, used for correcting the PU profile in simulation to the profile in data, is 4.6\%~\cite{Sirunyan:2018nqx}. The effect of this uncertainty on the signal normalization amounts to approximately 1\%.
\item 
Uncertainties in the efficiency of the photon plus $\PGt$-like jet trigger are 
2.2--3.4\% for the photon part and 5.3--5.6\% for the ditrack part.
\item Photon identification efficiencies are derived from $\PZ\to\Pe\Pe$ events and their uncertainty is propagated to the selection efficiency~\cite{CMS:2020uim}. The uncertainties are $\pt$ and $\eta$ dependent and are at most 1.5\%. 
\item The uncertainty in the tracking efficiency amounts to 2.3--2.4\% per track, corresponding to 4.6--4.8\% per ditrack, depending on the data-taking period. This uncertainty is determined by comparing ratios of \PDstz meson decay chains in data and simulation~\cite{CMS-DP-2022-012}.
\item Uncertainties in the muon and electron identification, isolation, and trigger efficiencies arise from the method used to obtain those efficiencies and from the limited size of the simulated samples used in these studies. They amount to less than 1.0\% (1.5\%) for muons (electrons)~\cite{Sirunyan:2018,CMS:2020uim}. 
\item Uncertainties affecting the charged and neutral meson isolation efficiencies are taken as the maximum difference between the efficiency as measured in data dimuon events and in simulated signal events. It ranges from 1.7\% to 2.8\%, depending on the decay channel and the type of isolation.
\item Uncertainties in jet energy scale and resolution corrections are only relevant for the VBF categories and are smaller than 3.5\%.
\item The theoretical QCD scale uncertainty is calculated by varying the renormalization and factorization scales. The maximum and minimum envelope of the resulting distributions are taken as the up/down variations. These uncertainties amount to 3.9\% for the \ggH production cross section. For the VBF, WH, ZH, and $\ttbar\PH$ production cross sections, the uncertainties are 0.4\%, 0.7\%, 3.8\%, and 2.6\%, respectively \cite{deFlorian:2016spz}.
\item Uncertainties from the choice of PDF set and the value of the strong coupling constant (\alpS) depend on the Higgs boson production mode and range from 1.6 to 3.2\%. These uncertainties are derived through reweighting the events with several alternative sets of weights. These weights are derived according to the prescriptions in PDF4LHC~\cite{Butterworth:2015oua} and NNPDF3.1~\cite{Ball:2017nwa}.
\item Parton shower modeling uncertainties arise from the renormalization scale of QCD-induced initial-state radiation (ISR) and FSR in \PYTHIA. Variations of renormalization parameters of the ISR and FSR consist of four
combinations obtained by keeping one constant and doubling or halving the other. The corresponding uncertainties are in the range 1.0--3.0\%.
\end{enumerate}

The largest impact on the UL is from the background shape uncertainties. Uncertainties in the signal yield from the acceptance effects in theory calculations are found to be negligible, as are those arising from the photon energy scale and resolution.

Theoretical uncertainties in the production cross sections, and the uncertainties due to the choice of PDF set and the value of \alpS are treated as correlated between the different data-taking periods. The uncertainty in the integrated luminosity measurement is treated as partially correlated between these periods. The other experimental uncertainties are treated as uncorrelated between the different data-taking periods. 

\section{Results}
\label{sec:results}
\begin{figure*}[htbp]
\centering
\includegraphics[width=0.4\textwidth]{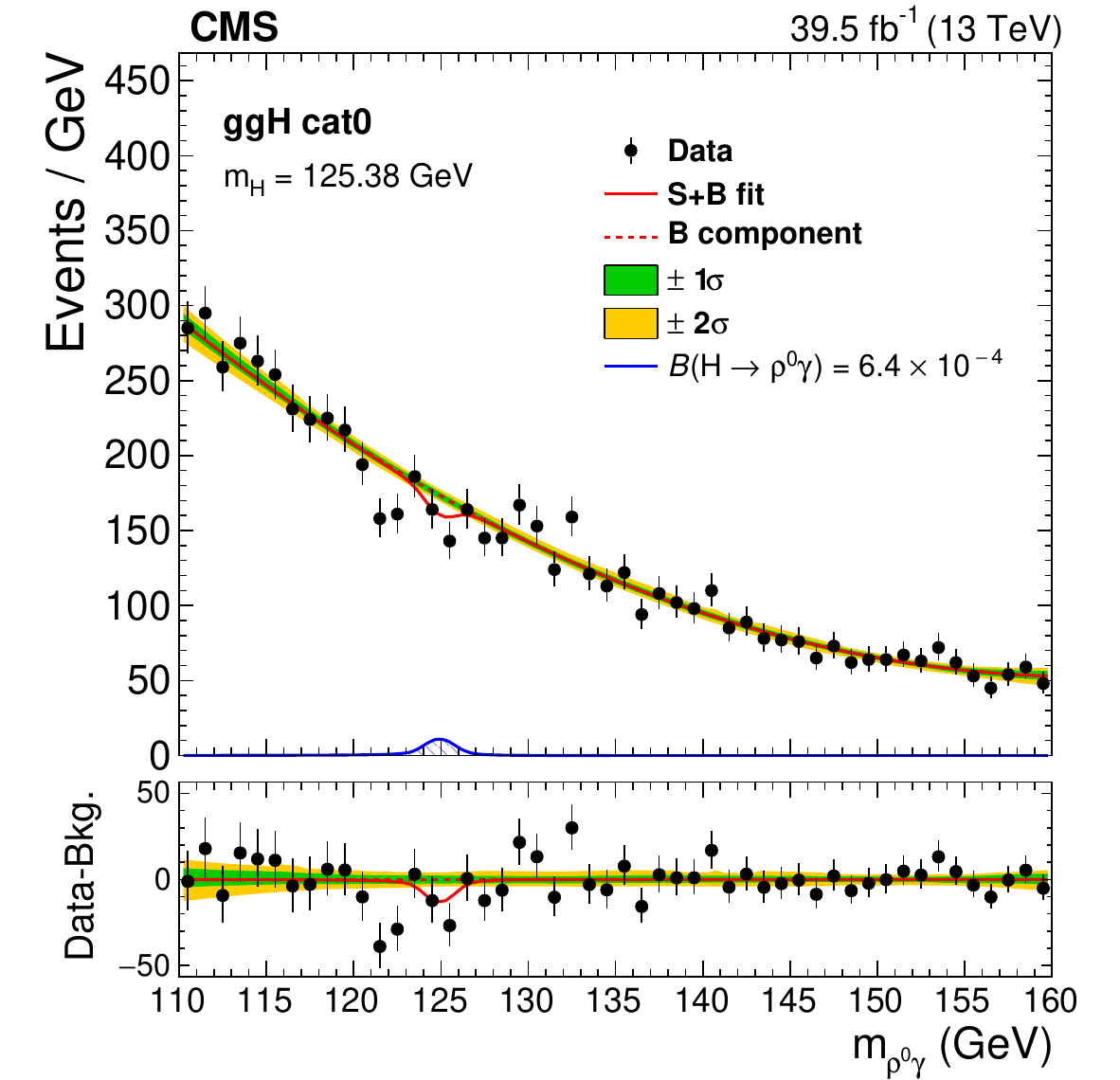}
\includegraphics[width=0.4\textwidth]{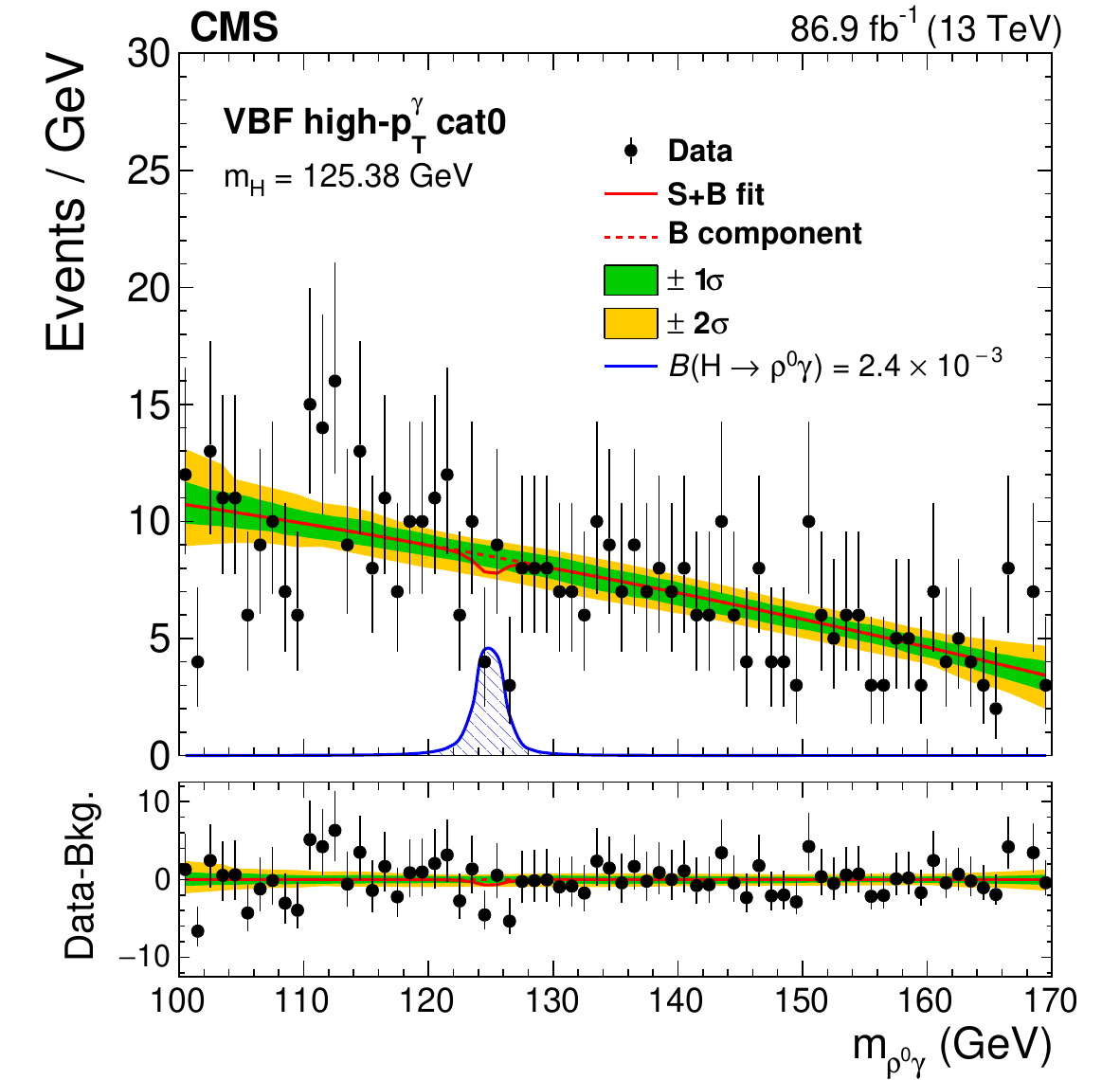}
\includegraphics[width=0.4\textwidth]{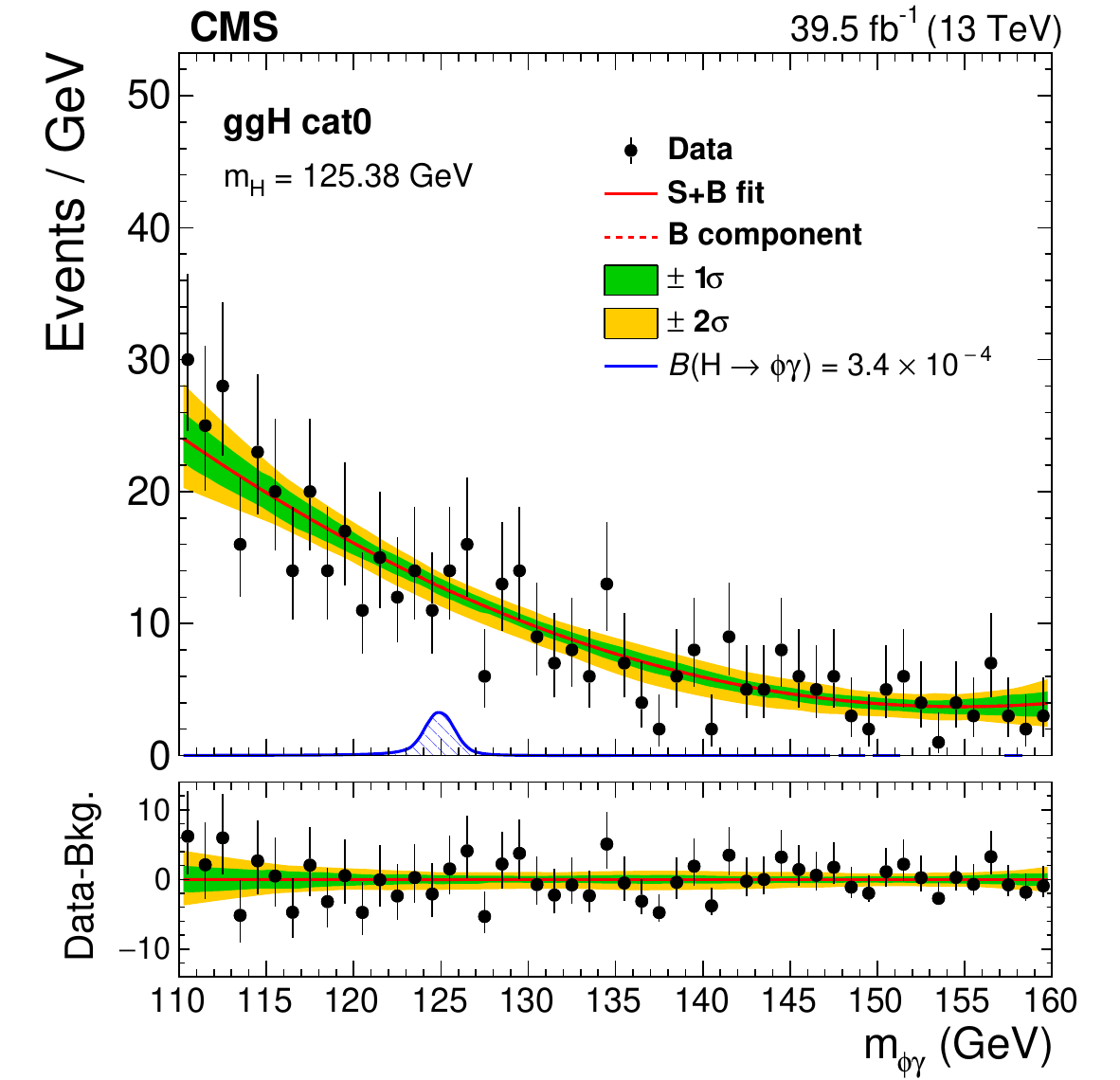}
\includegraphics[width=0.4\textwidth]{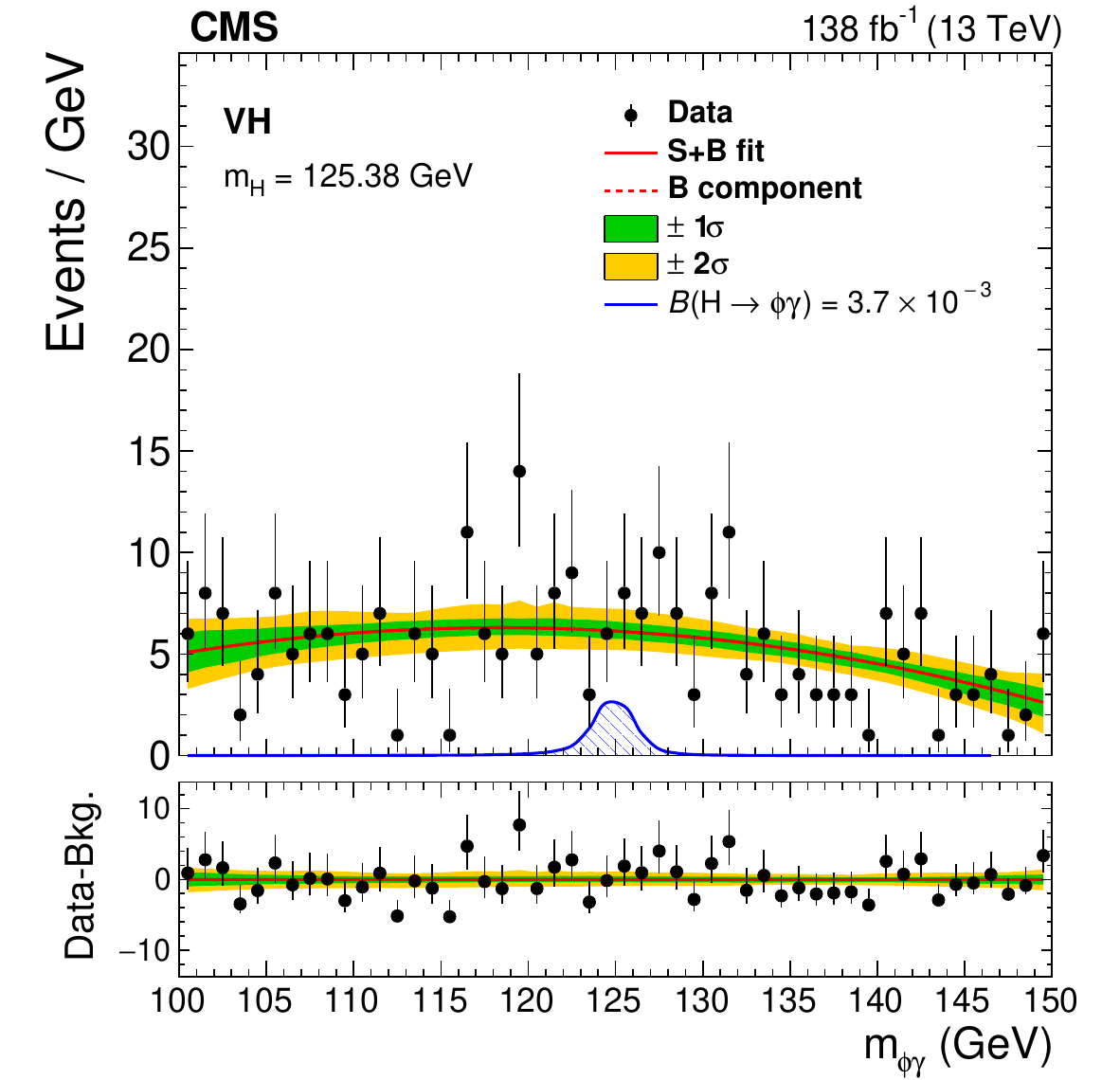}
\includegraphics[width=0.4\textwidth]{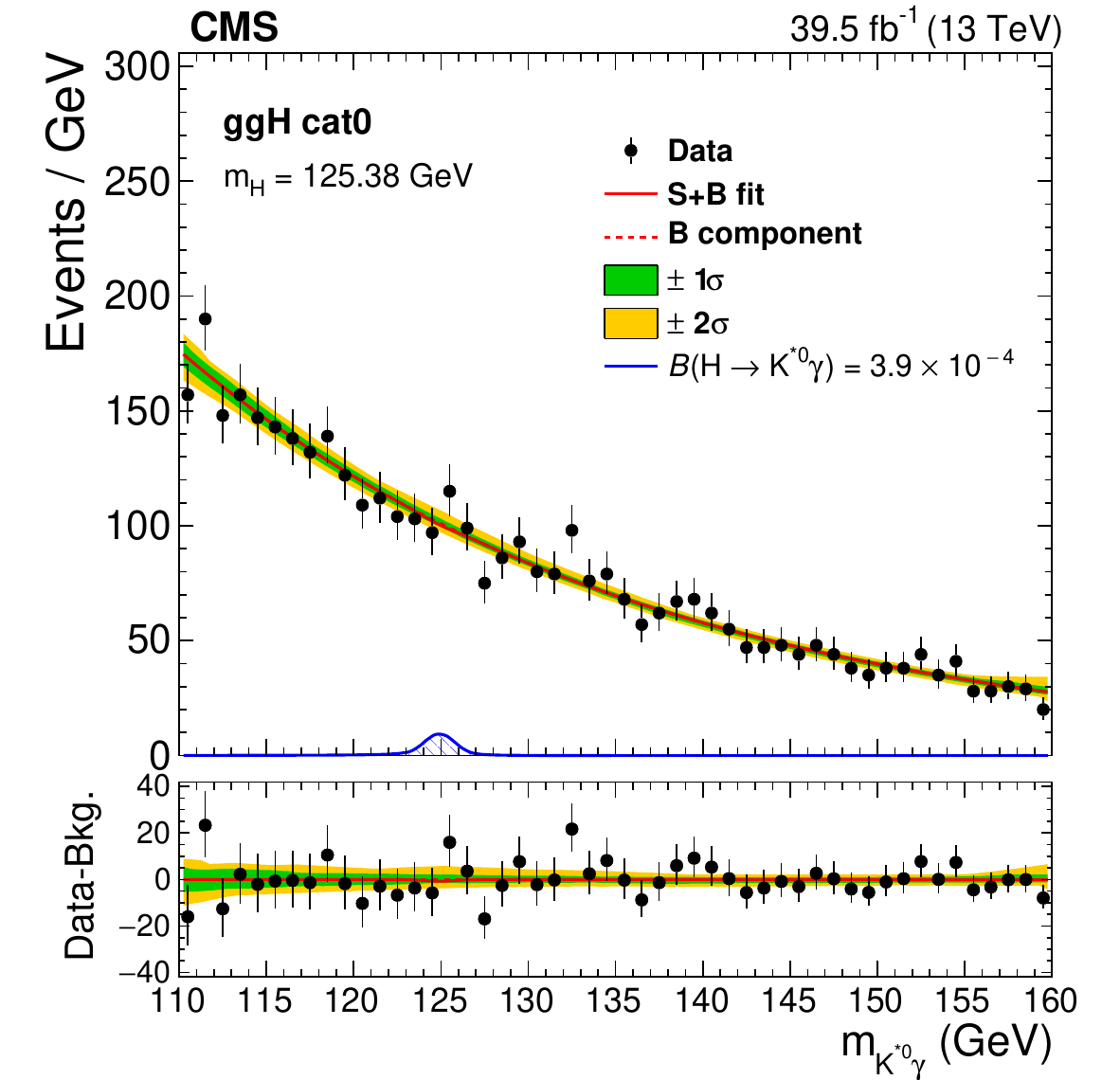}
\includegraphics[width=0.4\textwidth]{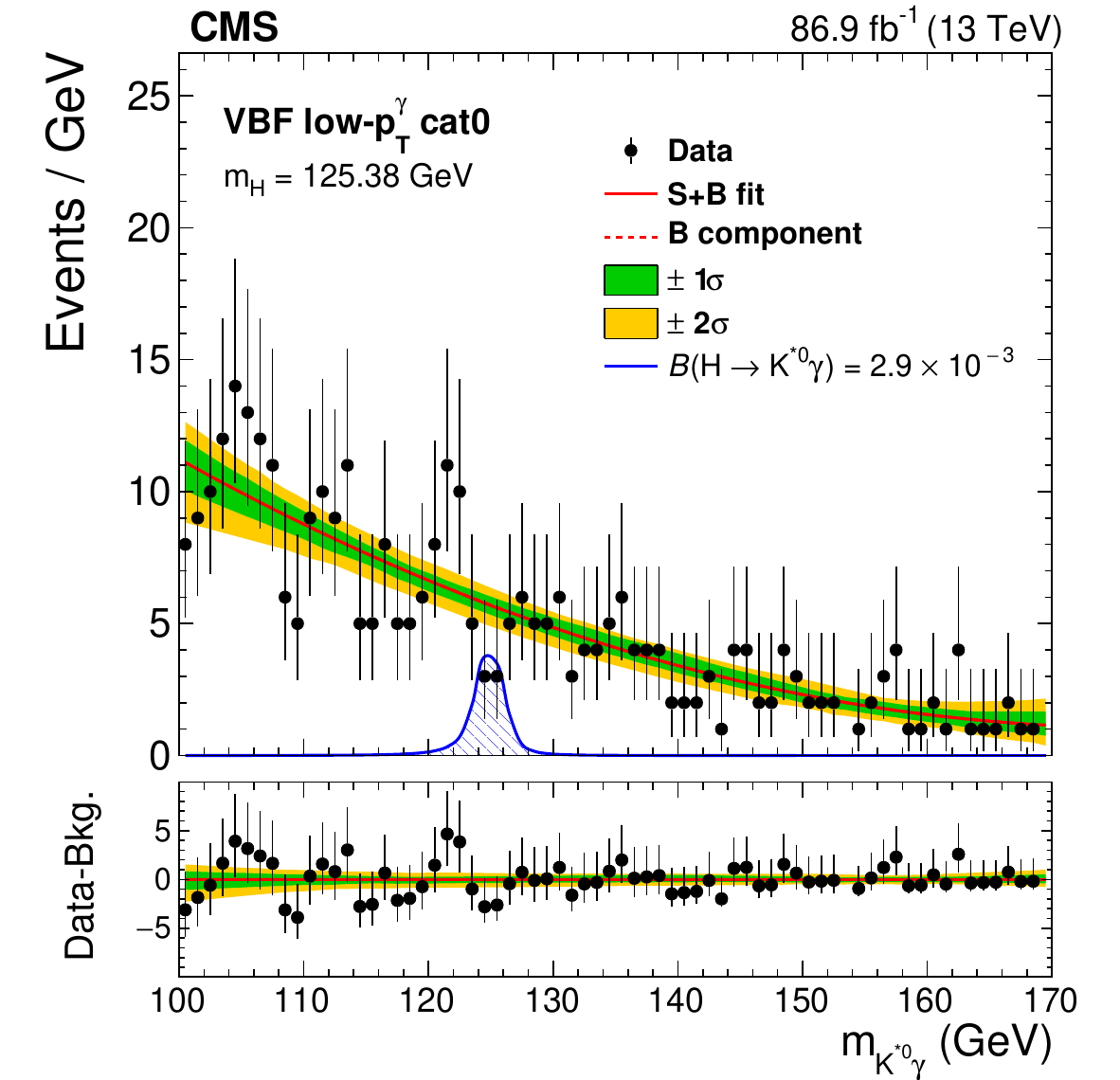}
\caption{Post-fit $\mllhad$ distributions in data and the background model for \ggH cat0 (upper left) and VBF high-$\pt^\PGg$ cat0 (upper right) of the $\Hgrho$ search, \ggH cat0 (middle left) and the $\PV\PH$ category (middle right) of the $\Hgphi$ search, \ggH cat0 (lower left) and the VBF low-$\pt^\PGg$ cat0 (lower right) of the $\Hgkstar$ search. 
The signal simulation is shown normalized at a branching fraction corresponding to the expected UL. Signal and background fit components, as well as 1 and 2 standard deviation(s) uncertainty bands are also shown.}
\label{fig:sigplots}
\end{figure*}

\begin{figure*}[htbp]                                                                                                                                                                                                                                                                             
\centering                                                                                                                                                                                                                                                                                        
\includegraphics[width=0.49\textwidth]{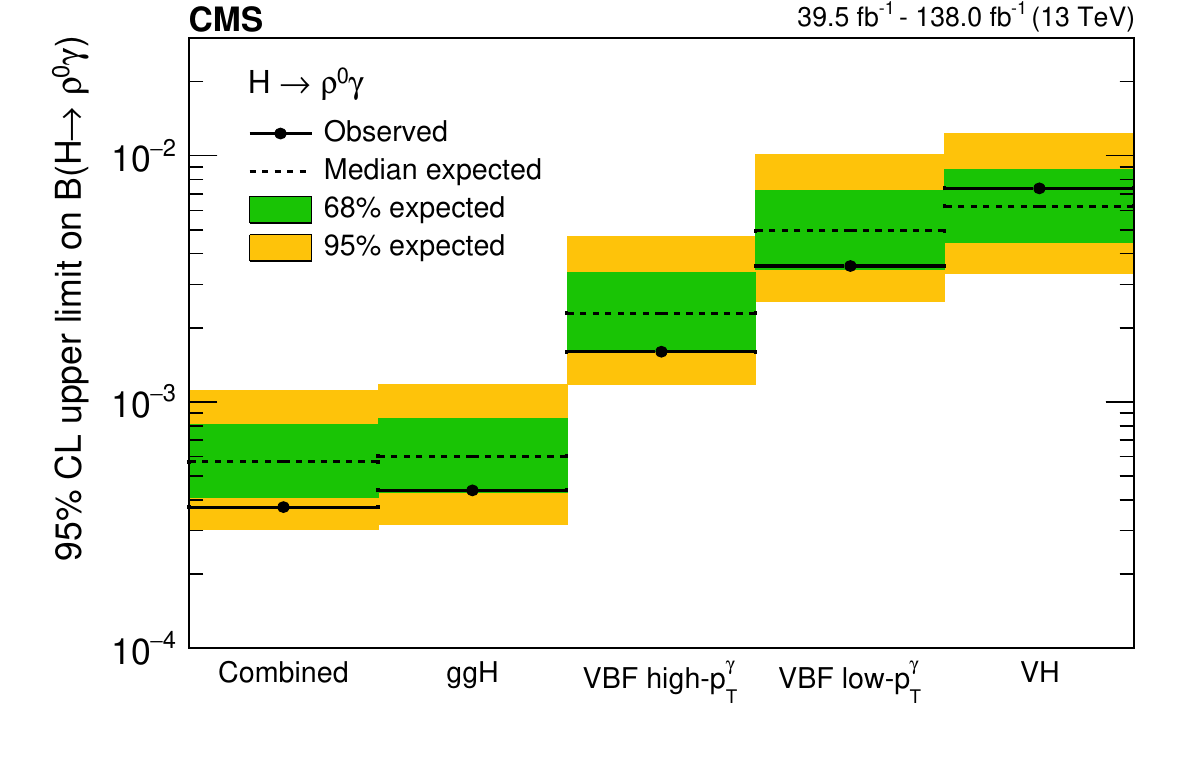}                                                                                                                                                                                                                                          
\includegraphics[width=0.49\textwidth]{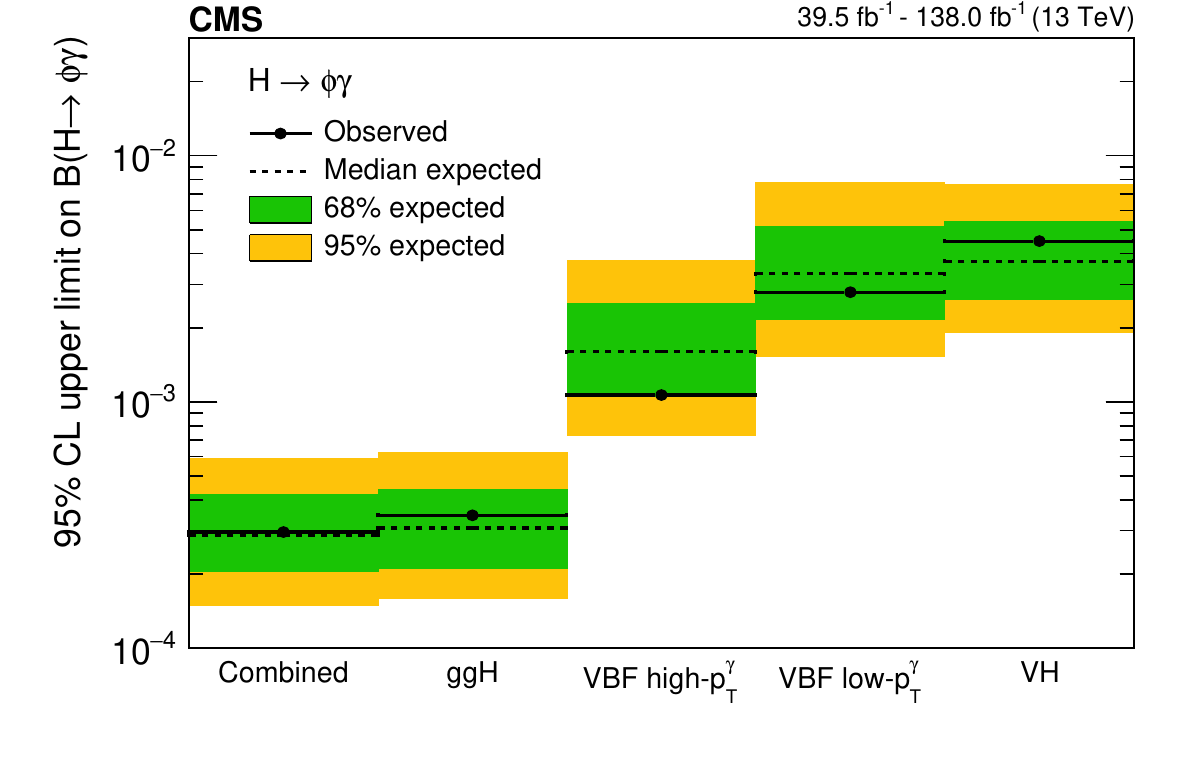} \\                                                                                                                                                                                                                                       
\includegraphics[width=0.49\textwidth]{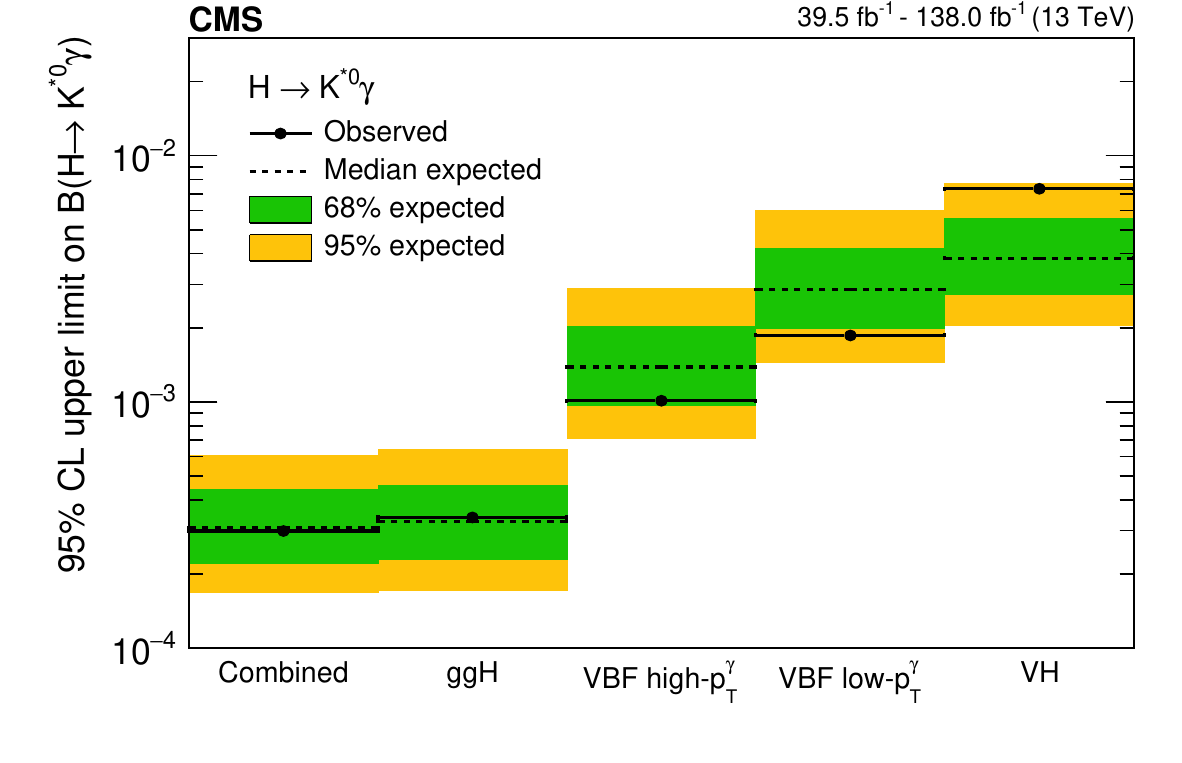}                                                                                                                                                                                                                                          
\caption{Expected and observed UL on $\mathcal{B}(\Hgrho)$ (upper left), $\mathcal{B}(\Hgphi)$ (upper right), and $\mathcal{B}(\Hgkstar)$ (lower) split by analysis categories (with subcategories combined, where applicable) and their overall combination. Green and yellow bands correspond to 68\% and 95\% confidence intervals on the expected upper limits.}

\label{fig:limits}                                                                                                                                                                                                                                                                                
\end{figure*}

The results of this analysis have been determined using the CMS statistical analysis tool \textsc{Combine}~\cite{CMS:2024onh}.
The $\mllhad$ distributions in data and the background estimations are shown in Fig.~\ref{fig:sigplots} for some selected analysis categories. The $\Hgrho$, $\Hgphi$, and $\Hgkstar$ signals are also shown, normalized to the UL of the branching fractions expected for the selected categories. No excess above the background expectation is observed. The results are presented as UL on $\mathcal{B}(\Hgrho)$, $\mathcal{B}(\Hgphi)$, and $\mathcal{B}(\Hgkstar)$ set at 95\% confidence level (\CL). Limits are set using the modified frequentist $\CLs$ approach~\cite{CLS2,Read:2002hq}, in which a profile-likelihood ratio modified for UL setting is used as the test statistic, making use of the asymptotic approximation~\cite{Cowan:2010js}. Systematic uncertainties are incorporated in the likelihood as nuisance parameters. Figure~\ref{fig:limits} reports the expected and observed 95\% \CL UL according to analysis category, where the two BDT-based sub-categories of \ggH and VBF have been merged. The \ggH category dominates the search sensitivity for all decay channels. By combining all the categories, the observed (expected) UL on $\mathcal{B}(\Hgrho)$, $\mathcal{B}(\Hgphi)$, and $\mathcal{B}(\Hgkstar)$ 
are $3.7 \times 10^{-4}$ ($5.7 \times 10^{-4}$), $3.0 \times 10^{-4}$ ($2.9 \times 10^{-4}$), and $3.0 \times 10^{-4}$ ($3.1 \times 10^{-4}$), respectively. These values correspond to 22 times the SM expectation for the $\Hgrho$ decay and 129 times the SM expectation for the $\Hgphi$ decay. For the strongly suppressed $\Hgkstar$ decay, this limit corresponds to $3.0 \times 10^{15}$ times the value expected in the SM. The observed and median expected exclusion limits for the branching fractions at 95\% \CL for the Higgs boson decays are listed in Table~\ref{tab:limits}.

\begin{table*}[htbp]
\footnotesize
  \centering
  \topcaption{Exclusion limits at 95\% \CL on the branching fractions of the Higgs boson decays. The limits are reported for the four categories (with subcategories combined, where applicable) and for their overall combination. Observed and median expected limits with the upper and lower bounds in the expected 68\% \CL intervals are also reported.}
  \renewcommand{\arraystretch}{1.3}
  \cmsTable{
    \begin{tabular}{lrrrrrr}
    \hline
       & \multicolumn{2}{c}{$\mathcal{B}(\Hgrho)$}  & \multicolumn{2}{c}{$\mathcal{B}(\Hgphi)$} &  \multicolumn{2}{c}{$\mathcal{B}(\Hgkstar)$}\\ [2ex]
         Category & Exp.$(\times10^{-4})$ & Obs.$(\times10^{-4})$ & Exp.$(\times10^{-4})$ & Obs.$(\times10^{-4})$ & Exp.$(\times10^{-4})$ & Obs.$(\times10^{-4})$\\
         \hline
        $\PV\PH$ & $62^{+26}_{-18}$ & 74 & $37^{+17}_{-11}$ & 45 & $38^{+17}_{-11}$ & 73 \\[2ex]
        Low-$\pt^\PGg$ VBF & $50^{+23}_{-15}$ & 36 & $33^{+19}_{-12}$ & 28 & $29^{+13}_{-9}$  & 19\\[2ex]
        High-$\pt^\PGg$ VBF & $23^{+11}_{-7}$ & 16 & $16^{+9}_{-6}$ & 11 & $14^{+7}_{-4}$ & 10 \\[2ex]
        \ggH &$6.0^{+2.5}_{-1.7}$ & 4.4 & $3.1^{+1.3}_{-1.0}$ & 3.5 & $3.3^{+1.3}_{-1.0}$ & 3.4\\[2ex]
        Combined & $5.7^{+2.4}_{-1.6}$ & 3.7 &$2.9^{+1.3}_{-0.8}$ & 3.0 & $3.1^{+1.3}_{-0.9}$ &3.0 \\[2ex]
        \hline
    \end{tabular}
    }
\label{tab:limits}  
\end{table*}

\section{Summary}
\label{sec:summary}
A search was presented for Higgs boson decays into a photon and a $\PGrzM$, $\Pgf$, or $\PKstzM$ meson, using pp collision data at $\sqrt{s}=13\TeV$ collected by the CMS experiment at the LHC. Events are selected assuming the mesons decay into a pair of charged pions, a pair of charged kaons, or a charged kaon and pion, respectively. Depending on the Higgs boson production mode, different signal triggering and reconstruction techniques are adopted. The analyzed data sets correspond to an integrated luminosity varying between 39.5 and 138\fbinv, depending on the targeted final state. After combining various data sets and categories, no significant excess above the background expectations is observed. Upper limits at the 95\% confidence level on the Higgs boson branching fractions into $\PGrzM\PGg$, $\Pgf\PGg$, and $\PKstzM\PGg$ are determined to be $3.7 \times 10^{-4}$, $3.0 \times 10^{-4}$, and $3.0 \times 10^{-4}$, respectively. Limits for the $\PGrzM\PGg$ and $\PGfM\PGg$ channels constitute the most stringent experimental limits to date. 

\begin{acknowledgments}
  We congratulate our colleagues in the CERN accelerator departments for the excellent performance of the LHC and thank the technical and administrative staffs at CERN and at other CMS institutes for their contributions to the success of the CMS effort. In addition, we gratefully acknowledge the computing centres and personnel of the Worldwide LHC Computing Grid and other centres for delivering so effectively the computing infrastructure essential to our analyses. Finally, we acknowledge the enduring support for the construction and operation of the LHC, the CMS detector, and the supporting computing infrastructure provided by the following funding agencies: SC (Armenia), BMBWF and FWF (Austria); FNRS and FWO (Belgium); CNPq, CAPES, FAPERJ, FAPERGS, and FAPESP (Brazil); MES and BNSF (Bulgaria); CERN; CAS, MoST, and NSFC (China); MINCIENCIAS (Colombia); MSES and CSF (Croatia); RIF (Cyprus); SENESCYT (Ecuador); ERC PRG, RVTT3 and MoER TK202 (Estonia); Academy of Finland, MEC, and HIP (Finland); CEA and CNRS/IN2P3 (France); SRNSF (Georgia); BMBF, DFG, and HGF (Germany); GSRI (Greece); NKFIH (Hungary); DAE and DST (India); IPM (Iran); SFI (Ireland); INFN (Italy); MSIP and NRF (Republic of Korea); MES (Latvia); LMTLT (Lithuania); MOE and UM (Malaysia); BUAP, CINVESTAV, CONACYT, LNS, SEP, and UASLP-FAI (Mexico); MOS (Montenegro); MBIE (New Zealand); PAEC (Pakistan); MES and NSC (Poland); FCT (Portugal);  MESTD (Serbia); MCIN/AEI and PCTI (Spain); MOSTR (Sri Lanka); Swiss Funding Agencies (Switzerland); MST (Taipei); MHESI and NSTDA (Thailand); TUBITAK and TENMAK (Turkey); NASU (Ukraine); STFC (United Kingdom); DOE and NSF (USA).
   
  \hyphenation{Rachada-pisek} Individuals have received support from the Marie-Curie programme and the European Research Council and Horizon 2020 Grant, contract Nos.\ 675440, 724704, 752730, 758316, 765710, 824093, 101115353, 101002207, and COST Action CA16108 (European Union); the Leventis Foundation; the Alfred P.\ Sloan Foundation; the Alexander von Humboldt Foundation; the Science Committee, project no. 22rl-037 (Armenia); the Belgian Federal Science Policy Office; the Fonds pour la Formation \`a la Recherche dans l'Industrie et dans l'Agriculture (FRIA-Belgium); the F.R.S.-FNRS and FWO (Belgium) under the ``Excellence of Science -- EOS" -- be.h project n.\ 30820817; the Beijing Municipal Science \& Technology Commission, No. Z191100007219010 and Fundamental Research Funds for the Central Universities (China); the Ministry of Education, Youth and Sports (MEYS) of the Czech Republic; the Shota Rustaveli National Science Foundation, grant FR-22-985 (Georgia); the Deutsche Forschungsgemeinschaft (DFG), among others, under Germany's Excellence Strategy -- EXC 2121 ``Quantum Universe" -- 390833306, and under project number 400140256 - GRK2497; the Hellenic Foundation for Research and Innovation (HFRI), Project Number 2288 (Greece); the Hungarian Academy of Sciences, the New National Excellence Program - \'UNKP, the NKFIH research grants K 131991, K 133046, K 138136, K 143460, K 143477, K 146913, K 146914, K 147048, 2020-2.2.1-ED-2021-00181, and TKP2021-NKTA-64 (Hungary); the Council of Science and Industrial Research, India; ICSC -- National Research Centre for High Performance Computing, Big Data and Quantum Computing and FAIR -- Future Artificial Intelligence Research, funded by the NextGenerationEU program (Italy); the Latvian Council of Science; the Ministry of Education and Science, project no. 2022/WK/14, and the National Science Center, contracts Opus 2021/41/B/ST2/01369 and 2021/43/B/ST2/01552 (Poland); the Funda\c{c}\~ao para a Ci\^encia e a Tecnologia, grant CEECIND/01334/2018 (Portugal); the National Priorities Research Program by Qatar National Research Fund; the Ministry of Science and Higher Education, projects no. FSWU-2023-0073 and no. FSWW-2020-0008 (Russia); MCIN/AEI/10.13039/501100011033, ERDF ``a way of making Europe", and the Programa Estatal de Fomento de la Investigaci{\'o}n Cient{\'i}fica y T{\'e}cnica de Excelencia Mar\'{\i}a de Maeztu, grant MDM-2017-0765 and Programa Severo Ochoa del Principado de Asturias (Spain); the Chulalongkorn Academic into Its 2nd Century Project Advancement Project, and the National Science, Research and Innovation Fund via the Program Management Unit for Human Resources \& Institutional Development, Research and Innovation, grant B39G670016 (Thailand); the Kavli Foundation; the Nvidia Corporation; the SuperMicro Corporation; the Welch Foundation, contract C-1845; and the Weston Havens Foundation (USA).
\end{acknowledgments}
\bibliography{auto_generated}

\cleardoublepage \appendix\section{The CMS Collaboration \label{app:collab}}\begin{sloppypar}\hyphenpenalty=5000\widowpenalty=500\clubpenalty=5000
\cmsinstitute{Yerevan Physics Institute, Yerevan, Armenia}
{\tolerance=6000
A.~Hayrapetyan, A.~Tumasyan\cmsAuthorMark{1}\cmsorcid{0009-0000-0684-6742}
\par}
\cmsinstitute{Institut f\"{u}r Hochenergiephysik, Vienna, Austria}
{\tolerance=6000
W.~Adam\cmsorcid{0000-0001-9099-4341}, J.W.~Andrejkovic, T.~Bergauer\cmsorcid{0000-0002-5786-0293}, S.~Chatterjee\cmsorcid{0000-0003-2660-0349}, K.~Damanakis\cmsorcid{0000-0001-5389-2872}, M.~Dragicevic\cmsorcid{0000-0003-1967-6783}, P.S.~Hussain\cmsorcid{0000-0002-4825-5278}, M.~Jeitler\cmsAuthorMark{2}\cmsorcid{0000-0002-5141-9560}, N.~Krammer\cmsorcid{0000-0002-0548-0985}, A.~Li\cmsorcid{0000-0002-4547-116X}, D.~Liko\cmsorcid{0000-0002-3380-473X}, I.~Mikulec\cmsorcid{0000-0003-0385-2746}, J.~Schieck\cmsAuthorMark{2}\cmsorcid{0000-0002-1058-8093}, R.~Sch\"{o}fbeck\cmsorcid{0000-0002-2332-8784}, D.~Schwarz\cmsorcid{0000-0002-3821-7331}, M.~Sonawane\cmsorcid{0000-0003-0510-7010}, W.~Waltenberger\cmsorcid{0000-0002-6215-7228}, C.-E.~Wulz\cmsAuthorMark{2}\cmsorcid{0000-0001-9226-5812}
% ...
\par}
\vskip\cmsinstskip
\dag:~Deceased\\
$^{1}$Also at Yerevan State University, Yerevan, Armenia\\
$^{2}$Also at TU Wien, Vienna, Austria\\
% ...
\end{sloppypar}
\end{document}